\documentclass[prb,superscriptaddress,twocolumn,floatfix,amsmath,amssymb,showpacs]{revtex4-1}

\usepackage{graphicx}
\usepackage{dcolumn}
\usepackage{bm}

\begin{document}

\title{Self doping effect and successive magnetic transitions in superconducting Sr$_2$VFeAsO$_3$}

\author{Guang-Han Cao}
\email[corresponding author: ]{ghcao@zju.edu.cn}
\affiliation{Department of Physics, Zhejiang University, Hangzhou
310027, China} \affiliation{State Key Lab of Silicon Materials,
Zhejiang University, Hangzhou 310027, China}

\author{Zhifeng Ma,$^1$ Cao Wang,$^1$ Yunlei Sun,$^1$ Jinke Bao,$^1$ Shuai Jiang,$^1$ Yongkang Luo,$^1$ Chunmu Feng,$^1$ Yi Zhou}
\affiliation{Department of Physics, Zhejiang University, Hangzhou
310027, China}

\author{Zhi Xie,$^3$ Fengchun Hu,$^3$ Shiqiang Wei}
\affiliation{National Synchrotron Radiation Laboratory, University
of Science and Technology of China, Hefei 230029, China}

\author{I. Nowik,$^4$ I. Felner}
\affiliation{Racah Institute of Physics, The Hebrew University,
Jerusalem 91904, Israel}

\author{Lei Zhang}
\affiliation{High Magnetic Field Laboratory, Chinese Academy of
Sciences, Hefei 230031, China}

\author{Zhu'an Xu}
\affiliation{Department of Physics, Zhejiang University, Hangzhou
310027, China} \affiliation{State Key Lab of Silicon Materials,
Zhejiang University, Hangzhou 310027, China}

\author{Fu-Chun Zhang}
\affiliation{Department of Physics, Zhejiang University, Hangzhou
310027, China} \affiliation{Department of Physics, The University of
Hong Kong, Hong Kong, China}

\date{\today}

\begin{abstract}
We have studied a quinary Fe-based superconductor Sr$_2$VFeAsO$_3$
by the measurements of x-ray diffraction, x-ray absorption,
M\"{o}ssbauer spectrum, resistivity, magnetization and specific
heat. This apparently undoped oxyarsenide is shown to be self doped
via electron transfer from the V$^{3+}$ ions. We observed successive
magnetic transitions within the VO$_2$ layers: an antiferromagnetic
transition at 150 K followed by a weak ferromagnetic transition at
55 K. The spin orderings within the VO$_2$ planes are discussed
based on mixed valence of V$^{3+}$ and V$^{4+}$.
\end{abstract}


\maketitle
\section{\label{sec:level1}Introduction}
The discovery of high-temperature superconductivity (HTSC) in
layered Fe-based compounds\cite{hosono,review} represents an
important breakthrough in the field of condensed matter physics. So
far, dozens of new Fe-based superconductors in several
crystallographic types have been discovered.\cite{review} The common
structural unit for the HTSC is Fe$_2$$X_2$ ($X$=As, Te, etc.)
layers. The undoped Fe$_2$$X_2$ layers with formally divalent iron
usually exhibit antiferromagnetic (AFM) spin-density-wave (SDW)
instability.\cite{dpc,bw} HTSC emerges as the SDW ordering is
suppressed by various kinds of
doping\cite{hosono,ren-oxygen,wen-hole,wang-Th,Co,Ni,ren-P} or
applying pressures\cite{pressure}. Both the magnetic ordering and
superconductivity were suggested to be related to the nesting
between the two cylinder-like Fermi surfaces near $\Gamma$ and M
points.\cite{singh,wnl,mazin1,kuroki,ding}

Recently, superconductivity at 37 K was reported in a quinary
oxyarsenide Sr$_2$VFeAsO$_3$ (hereafter called V21113),\cite{wen1}
consisting of perovskite-like Sr$_4$V$_2$O$_6$ and
anti-fluorite-type Fe$_2$As$_2$ block layers. The observed
superconductivity \emph{without extrinsic doping} challenges the
above paradigm, and immediately aroused several first principles
calculations.\cite{shein,pickett,wang,mazin2,nakamura} However, the
occurrence of superconductivity in the apparently undoped V21113 has
not been clarified. Moreover, the calculated property of the
Sr$_4$V$_2$O$_6$ layers is very much scattered: including magnetic
half-metallic,\cite{shein} nonmagnetic metallic,\cite{pickett,wang}
and magnetic (Mott-like) insulating\cite{nakamura} states.  Various
kinds of possibilities on the magnetic ground state for V21113 were
evaluated, but no conclusive result was given.\cite{mazin2}
Experimentally, few works\cite{wen1,wen2} have been devoted to this
new system. It is simply not clear whether the Fe-site AFM order
survives, let alone the magnetic property of the Sr$_4$V$_2$O$_6$
layers. The valence state of V, which is very crucial to understand
the appearance of superconductivity, remains a puzzle. X-ray
photoelectron spectroscopy (XPS) measurement suggested an
"unexpected" V$^{5+}$ even in the oxygen-deficient V21113 samples,
however, the result was then questioned because of the
surface-sensitive nature for the XPS method.\cite{wen2}

In this article, we report systematic experimental studies on the
structure and physical properties of V21113. We employed x-ray
absorption spectroscopy, capable of detecting the bulk properties,
to examine the valence state of V. Our data unambiguously reveal
mixed valence of V$^{3+}$ and V$^{4+}$, indicating that the
Fe$_2$As$_2$ layers are actually self doped due to the electron
charge transfer from V to Fe. Our measurements have also identified
two magnetic transitions at 55 K and 150 K, respectively, associated
with the VO$_2$ layers. The observed weak ferromagnetism below 55 K
and its coexistence with superconductivity at low temperatures make
the V21113 system very unique in the family of Fe-based
superconductors.

\section{\label{sec:level2}Experimental}
Synthesis of single-phase 21113 samples turned out to be
difficult,\cite{wen1} because of the tendency of
non-stoichiometry\cite{tegel} as well as the multi-interphase
solid-state reactions. We have improved the sample quality by
carefully controlling the stoichiometry. Samples of
Sr$_2$V$_{1-x}$Mg$_{x}$FeAsO$_3$ ($x$=0 corresponds to undoped
V21113, and $x$=0.1 refers to Mg-doped sample) were prepared using
fine powders ($\sim$ 200 mesh, purity $\geq$99.9 \%) of SrO, MgO, V,
Fe, Fe$_{2}$O$_{3}$ and As.  The accurately-weighed stoichiometric
mixture was loaded in an alumina tube, preventing possible reactions
with the quartz tube. The alumina tube, placed in a sealed quartz
ampoule, was heated slowly to 1273 K, holding for 24 h. After the
first time sintering, the sample was thoroughly ground, pressed into
pellets, and sintered again in vacuum ($< 10^{-4}$ Pa) at 1353 K for
48 h. All the operations of weighing, mixing, grinding, pelletizing,
etc. were carried out in an argon-filled glove-box with the water
and oxygen contents less than 0.1 ppm.

Several measurements were performed as follows. Powder x-ray
diffraction (XRD) was carried out using a D/Max-rA diffractometer
with Cu $K_{\alpha}$ radiation and a graphite monochromator. The
crystal structure at room temperature was refined by Rietveld
analysis. The V $K$-edge x-ray absorption spectra (XAS) were
measured in transmission mode using powdered samples on beam lines
U7B and U7C at the National Synchrotron Radiation Facility (NSRF),
Hefei, China. The water-cooled Si (111) plane bicrystal
monochromator was used. Calibration of spectrometer was made using
vanadium metal foil. The raw data were normalized according to the
previous literature.\cite{xas-v1} M\"{o}ssbauer spectra were
collected using a conventional constant acceleration drive. The
sources were 50 mCi $^{57}$Co:Rh. The velocity calibration was
performed with an $\alpha$-iron foil at room temperature. The
reported isomer shift (IS) values for iron are relative to the Fe
foil. The resistivity was measured on bar-shaped samples using a
standard four-terminal method. We employed Quantum Design PPMS-9 and
MPMS-5 to measure the specific heat and dc magnetization,
respectively.

\section{\label{sec:level3}Results and discussion}

The XRD patterns are shown in Fig. 1(a). In comparison with the
previous report\cite{wen1}, the intensity of impurity peaks
(relative to that of the main peak) at 2$\theta\sim$32$^{\circ}$ is
greatly reduced,  indicating that the sample's quality is
significantly improved. According to the multi-phase Rietveld
refinement\cite{note}, the amount of impurity phases Sr$_2$VO$_4$
and FeAs are less than 5\%. The refined structural parameters for
Sr$_2$V$_{1-x}$Mg$_{x}$FeAsO$_3$ ($x$=0 and 0.1) are listed in Table
I. The lattice constants of undoped V21113 are a little larger than
the reported values\cite{wen1}. The Mg substitution leads to a
lattice elongation. The As$-$Fe$-$As angle is very close to the
ideal value for a regular tetrahedron. On the other hand, the
V$-$O2$-$V angle obviously deviates from 180$^{\circ}$.

\begin{figure}
\includegraphics[width=7.5cm]{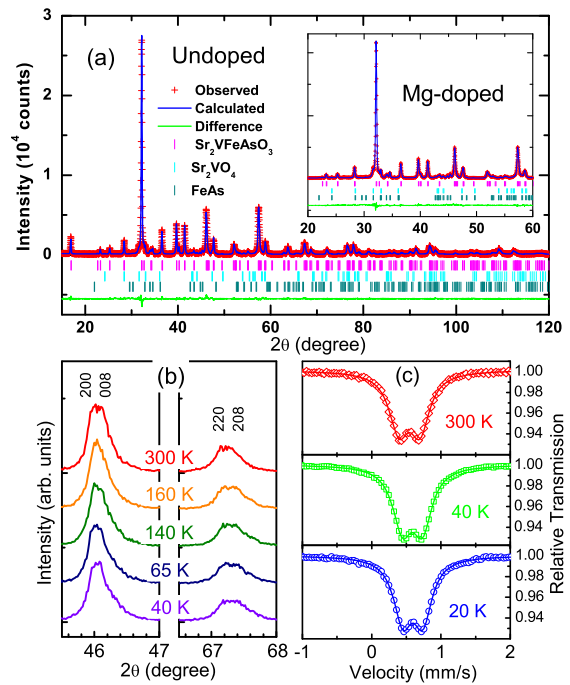}
\caption{(color online). (a) Rietveld refinement profiles for the
x-ray diffraction (XRD) data of Sr$_2$V$_{1-x}$Mg$_{x}$FeAsO$_3$
[$x$=0 and 0.1 (inset)]. (b) Selected XRD peaks at various
temperatures. (c) $^{57}$Fe M\"{o}ssbauer spectra at different
temperatures.}
\end{figure}

\begin{table}
\caption{\label{tab:table1}Crystallographic data of
Sr$_2$V$_{1-x}$Mg$_{x}$FeAsO$_3$ ($x$=0 and 0.1) at room
temperature. The space group is P$4/nmm$. The atomic coordinates are
as follows: Sr1 (0.25, 0.25, $z$); Sr2 (0.25, 0.25, $z$); V (0.25,
0.25, $z$); Fe (0.25, 0.75, 0); As (0.25, 0.25, $z$); O1 (0.25,
0.25, $z$); O2 (0.25, 0.75, $z$). BVS refers to bond valence
sum,\cite{bvs1} reflecting the formal valence of an ion
investigated.}
\begin{ruledtabular}
\begin{tabular}{lcr}
Compounds&$x$=0&$x$=0.1\\
\hline
$a$ ({\AA}) & 3.9352(1) &3.9334(2)\\
$c$ ({\AA}) & 15.6823(5) &15.7488(7)\\
$V$ ({\AA}$^3$) & 242.86(1) & 243.66(2)\\
$R_{wp}$&8.59 & 8.82\\
$S$&1.68 & 1.70\\
$z$ of Sr1&0.8090(1)& 0.8091(1)\\
$z$ of Sr2&0.5857(1)& 0.5865(1)\\
$z$ of V&0.3080(2) & 0.3074(2)\\
$z$ of As&0.0893(1) & 0.0880(2)\\
$z$ of O1&0.4248(4) & 0.4257(5)\\
$z$ of O2&0.2934(3) & 0.2952(4)\\

BVS$_\text{V}$ for V$^{\text{III}}$ & 2.95 &N.V.\\
BVS$_\text{V}$ for V$^{\text{IV}}$ & 3.29 &N.V.\\
BVS$_\text{V}$ for V$^{\text{V}}$ & 3.46 &N.V.\\
V$-$O2$-$V angle ($^{\circ}$) & 166.7(2) &168.8(2)\\
As$-$Fe$-$As angle ($^{\circ}$) & 109.1(1) &109.7(1)\\
\end{tabular}
\end{ruledtabular}
\end{table}

The low-temperature XRD experiments indicate no evidence of
structural phase transition in the whole temperatures measured. Some
of the data are shown in Fig. 1(b). First, the (200) peak does not
separate within the experimental limit, suggesting no usual
tetragonal-to-orthorhombic transition. Second, unlike the case in
$R$FeAsO ($R$=La, Sm, Gd and Tb),\cite{mcguire,luo} no obvious
splitting for the (220) peak can be seen. Thus, the
SDW-ordering-related structural phase transition\cite{luo} can be
ruled out in V21113. While the lattice parameters were found to
decrease with decreasing temperature, the precise temperature
dependence is expected to be measured by a future XRD study with
synchrotron radiations.

The M\"{o}ssbauer spectra [Fig. 1(c)] show a quadrupole doublet at
all the temperatures, in contrast with the sextet line for the AFM
SDW ordered state.\cite{mcguire,felner} From the data we deduced no
effective paramagnetic moment, therefore, the iron does not carry
magnetic moment on its own. Accordingly no magnetic ordering at the
Fe sublattice is expected. The IS value is similar to those observed
for other ferroarsenide systems,\cite{felner} suggesting that the
iron is basically divalent.

Fig. 2 shows the V $K$-edge XAS for V21113 and its reference
vanadium oxides. It has been well established that, in a number of V
compounds, the position of the characteristic absorptions (such as
absorption threshold, pre-edge peak, and main absorption edge)
varies linearly with the valence of V ($V_\text{V}$).\cite{xas-v1}
The main edge of V21113 lies in between those of V$_2$O$_3$ and
VO$_2$, clearly indicating the mixed valence state for V. The
pre-edge structure (to the left of the main edge) of V21113
resembles that of V$_2$O$_3$, suggesting that V$^{3+}$ is dominant.
One can fit the pre-edge of V21113 by the addition of weighed XAS of
V$_2$O$_3$ and VO$_2$. The best fitting gives $V_\text{V}$=3.22. In
fact, $V_\text{V}$ can be better determined by using the
characteristic absorption threshold ($E_{\text{thres}}$). As shown
in the insets of Fig. 2, $E_{\text{thres}}$ increases steadily on
going along V$_2$O$_3$$\rightarrow$ VO$_2$$\rightarrow$V$_2$O$_5$.
Assuming a linear relation of $E_{\text{thres}}$ vs $V_\text{V}$, we
obtain $V_\text{V} = 3.18 \pm 0.05$, consistent with the former
estimation. Here we note that the structural parameters of V21113
also support the conclusion about the V valence. The calculated
bond-valence-sum\cite{bvs1} values for V in Table I verify the
dominant V$^{3+}$ with minority of V$^{4+}$ in V21113.

\begin{figure}
\includegraphics[width=7.5cm]{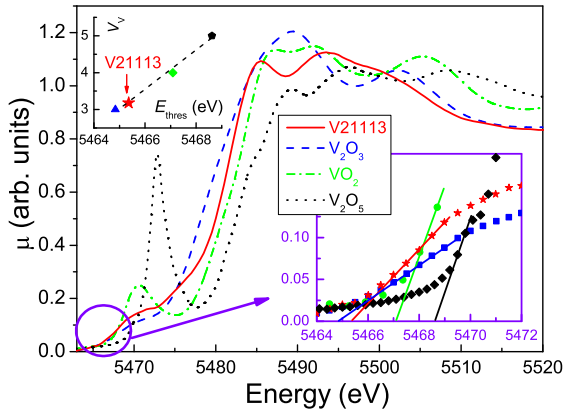}
\caption{\label{fig:epsart} (color online). Vanadium $K$-edge x-ray
absorption spectra for Sr$_2$VFeAsO$_3$ as well as some reference
compounds V$_2$O$_3$, VO$_2$ and V$_2$O$_5$. The insets show the
determination of the valence of V ($V_\text{V}$) by the absorption
threshold, $E_{\text{thres}}$.}
\end{figure}

Because of the overall charge neutrality, the mixed valence of V
immediately suggests that the Fe$_2$As$_2$ layers are "self" (or
internally) electron-doped by the charge transfer from the V$^{3+}$
ions (conversely, the Sr$_4$V$_2$O$_6$ layers are self hole-doped).
This conclusion is consistent with the measurements of Hall and
Seebeck coefficients, which show electron-dominant transport
behavior (not shown here), and also with the band
calculations\cite{pickett} which indicate the occupation number of
3$d$ electrons per V being less than 2. The absence of Fe-site AFM
ordering and appearance of SC transition (see below) are thus
naturally understood within the established paradigm for Fe-based
HTSC.

In Fig. 3, we plot the temperature dependence of resistivity
($\rho$), magnetic susceptibility ($\chi$) and specific heat ($C$)
for V21113 along with Sr$_2$V$_{0.9}$Mg$_{0.1}$FeAsO$_3$ as a
reference. The V21113 sample shows superconducting (SC) transition
at 24 K, which is somewhat lower than previously
reported\cite{wen1}. The lowered $T_c$ is mainly ascribed to the
"overdoped" state, since the self-doping level achieves $\sim$18\%
(from the obtained $V_\text{V}$ above). As we see, the Mg-doped
sample shows an enhanced $T_c$ of 30 K. This result does suggest
that the undoped V21113 be overdoped, because the Mg-for-V
substitution induces holes, which partially compensates the heavy
electron doping. Another reason for the lowered $T_c$ could be due
to the V-for-Fe anti-site occupation, as suggested by a very recent
report.\cite{V/Fe}

\begin{figure}
\includegraphics[width=7cm]{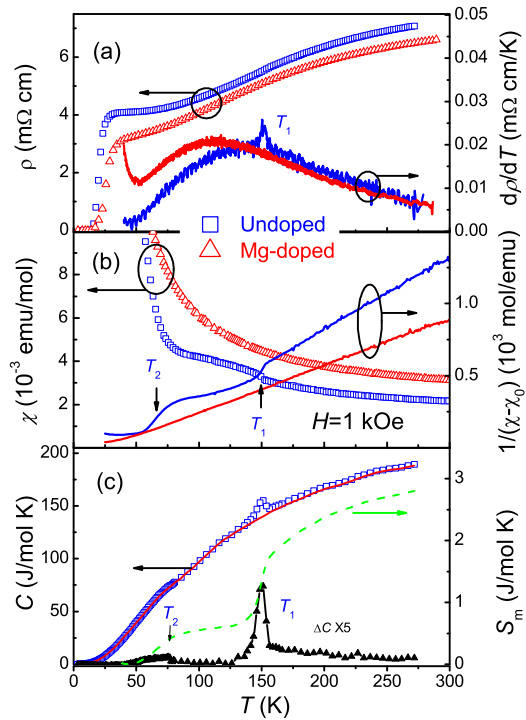}
\caption{(color online). Temperature dependence of resistivity (a),
magnetic susceptibility (b), and specific heat (c) for
Sr$_2$VFeAsO$_3$ and Sr$_2$V$_{0.9}$Mg$_{0.1}$FeAsO$_3$. The left
axes show the results after different data processing. $\chi_0$ is
1.4$\times10^{-3}$ and 2.0$\times10^{-3}$ emu/mol for the undoped
and Mg-doped samples, respectively.}
\end{figure}

For V21113, an anomaly at $T_1=$150 K is clearly shown in the
derivative of $\rho(T)$ as well as $\chi(T)$ and $C(T)$ curves. In
particular, the anomaly in $C(T)$ points to an intrinsic
second-order phase transition, since specific heat measures bulk
properties (note that the amount of impurities is less than 5\%).
For the Mg-doped sample, however, such an anomaly is completely
suppressed. Since the Fe-site magnetic ordering was ruled out by the
above M\"{o}ssbauer study, we conclude that the 150-K transition
V21113 is intrinsic and directly related to the V atoms.

The $\chi(T)$ data in the range of 300 K$>T>150$K obey the
Curie-Weiss law,
$\chi=\chi_{0}+C_{\text{Curie}}/(T-\theta_{\text{W}})$, where
$C_{\text{Curie}}$ denotes the Curie constant and
$\theta_{\text{W}}$ the Weiss temperature. Since the Fe atoms have
no effective paramagnetic moments, as revealed by the above
M\"{o}ssbauer study, we can estimate the effective magnetic moments
of V, which gives $P_{\text{eff}}$=1.2 $\mu_{B}$/V by the data
fitting. The significantly lowered $P_{\text{eff}}$ (compared with
the spin-only value for either V$^{3+}$ or V$^{4+}$) could be due to
the $d-p$ hybridization in connection with the mixed valence. The
fitted $\theta_{\text{W}}$ value is 52 K, implying dominant
ferromagnetic (FM) correlations between the magnetic moments. By
subtracting $C(T)$ of Mg-doped V21113 from that of the undoped
V21113, the magnetic entropy for the transition is calculated to be
$\sim$1.1 J$\cdot$mol$^{-1}\cdot$K$^{-1}$, which is much lower than
the expected value of $R$ln$(2S+1)$ ($S$=1 for V$^{3+}$; and $S$=1/2
for V$^{4+}$). The result suggests that short-range magnetic order
is established at the temperatures far above 150 K. Note that the
step-like increase in $\chi$ at the transition does not mean a FM
transition, because the $M(H)$ curve for 70 K$<T<$150 K is linear
[see Fig. 4(c)]. In fact, the AFM ordering of $A$ or $C$ type
possibly exhibits an increase in $\chi$ at the magnetic
transition.\cite{RVO3}

In addition to the 150-K transition, another anomaly at $T_2\sim$55
K can be clearly seen in 1/($\chi-\chi_0$), as shown in Fig. 3(b).
Specific heat data also show a small anomaly around 55 K. However,
the Mg-doped sample, which contains a little more impurities,
\emph{simultaneously} loses the two anomalies. This fact strongly
suggests that the 55-K anomaly is also intrinsic, and related to the
Sr$_4$V$_2$O$_6$ layers. Fig. 4 supplies more information about this
second transition. At low fields, the magnetization diverges at 55 K
for field-cooling (FC) and zero-field-cooling (ZFC) modes,
suggesting a ferromagnetic transition. The FM state is further
supported by the $M(H)$ curve, where a magnetic hysteresis loop is
clearly observed. The very small extrapolated residual magnetization
($M_0$$\sim$10$^{-3} \mu_{B}/$V at 35 K) and lack of saturation in
$M$ at high fields indicate that it is actually a weak
ferromagnetism (WFM). The WFM nature is consistent with the above
M\"{o}ssbauer spectra which show negligibly small transferred
hyperfine field (below 0.07 T) at the Fe site for $T$=20 and 40 K.
Nevertheless, the SC magnetic loop [Fig. 4(d)] exhibits obvious
asymmetry relative to the horizontal axis, suggesting significant
contributions of WFM in the SC state. It is noted that the
coexistence of SC and WFM differs with the previous finding of the
coexistence of SC and strong ferromagnetism (due to 4$f$ local
moments) in EuFe$_{2}$(As$_{0.7}$P$_{0.3}$)$_{2}$.\cite{ren-P}

\begin{figure}
\includegraphics[width=8cm]{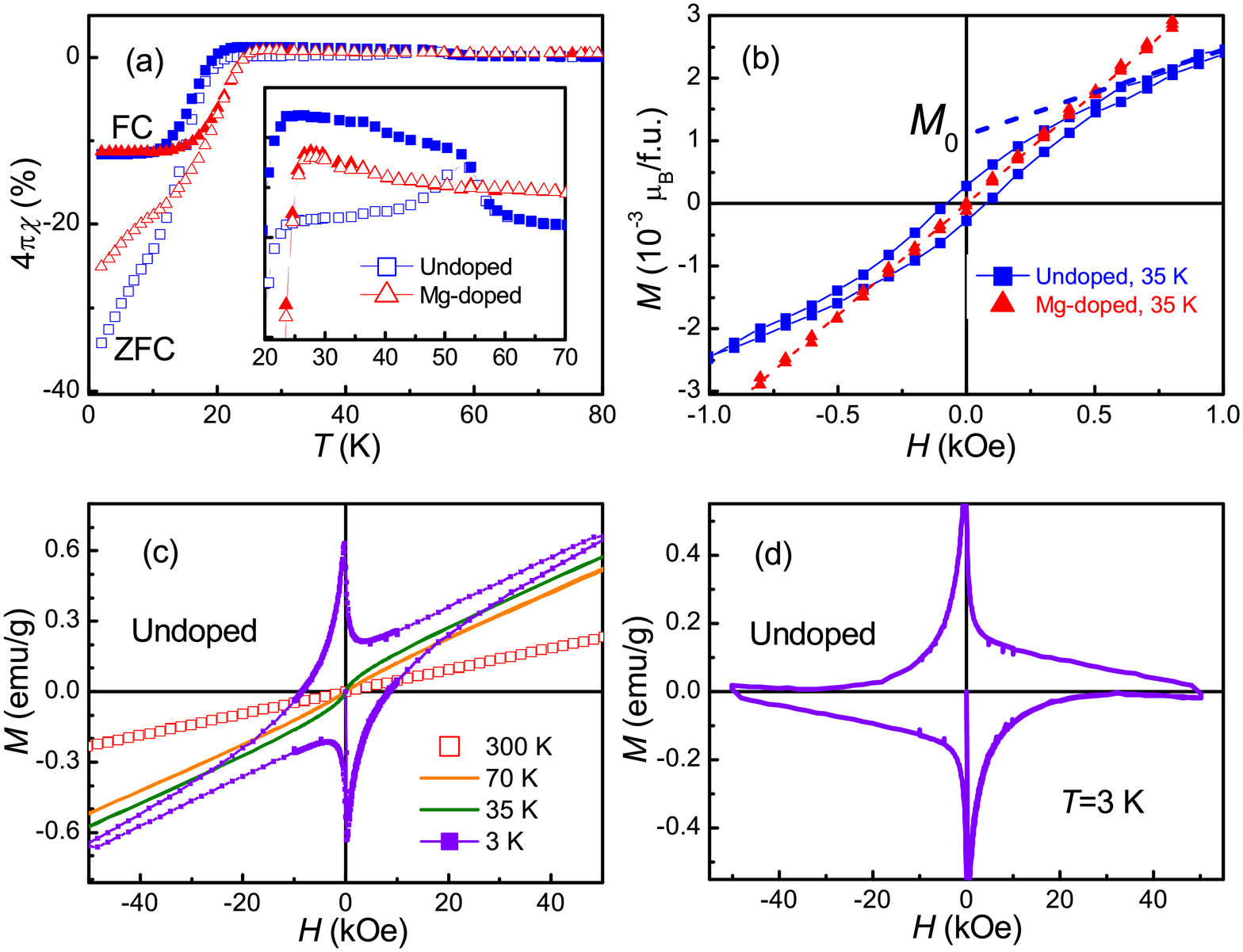}
\caption{(color online). Low-field $\chi(T)$ curves (a) and
low-field $M(H)$ curves (b) for Sr$_2$VFeAsO$_3$ (in comparison with
the Mg-doped sample). Panel (c) shows a full-range $M(H)$ curves.
Panel (d) displays the superconducting magnetization curve after
subtracting a paramagnetic background.}
\end{figure}

Now, let us discuss the possible origin of the two magnetic
transitions within the Sr$_4$V$_2$O$_6$ layers. The electronic
configurations of V$^{3+}$ and V$^{4+}$ are $d^2$ and $d^1$,
respectively. Within an ionic model, the lowest V $3d$ orbitals are
degenerate $d_{yz}$ and $d_{xz}$ due to the crystal field splitting.
In the case of pure V$^{3+}$, the electron occupation is
$d_{yz}^{\uparrow}$$d_{xz}^{\uparrow}$ according to the Hund's rule.
The ground state of the square lattice of V in the VO$_2$ planes is
AFM due to superexchange interaction.\cite{RVO3} In V21113, there
are $\sim$18\% V$^{4+}$ ions, and the Zener's double exchange favors
ferromagnetic spin arrangement in the same VO$_2$ plane for the gain
of kinetic energy. This could result in an in-plane FM state.
However, the electron hopping integrals in VO$_2$ planes are
direction-dependent. The amplitude of the hopping integral of
$d_{xz}$ along $y$-axis (or $d_{yz}$ along $x$-axis), $t'$, is much
smaller than that of $d_{xz}$ along $x$-axis (or $d_{yz}$ along
$y$-axis), $t$. This leads to another possible spin array in the
VO$_2$ planes, i.e. a $C$-type AFM state where the spins are
parallel along $x$-axis, and antiparallel along $y$-axis, or vice
versa. By a simplified mean field analysis, we estimate that the FM
state is more stable if $|t'|/J
> 2(1-\delta)/\delta$, with $J$ the neighboring spin coupling and
$\delta$ the concentration of V$^{4+}$.

The scenario of in-plane FM state coincides with the afore-mentioned
positive Weiss temperature. Thus the observed transition at
$T_1$=150 K may be ascribed to an $A$-type AFM ordering with AFM
coupling along the $c$-axis. The WFM transition at $T_2$=55 K may be
interpreted as the spin canting within the framework of double
exchange proposed by de Gennes,\cite{gennes} although usually one
would expect a more substantial magnetic moment. As for the scenario
of in-plane $C$-type AFM, the WFM transition is still an open
question. One possibility is due to the nonzero Dzyaloshinsky-Moriya
interaction, which plays an important role for the intriguing
magnetic responses in other vanadium oxides.\cite{YVO3}

\section{\label{sec:level4}Concluding remarks}
To summarize, we have demonstrated experimentally that
superconductivity in undoped Sr$_2$VFeAsO$_3$ is induced by a novel
self doping mechanism, i.e. an interlayer charge transfer from V to
Fe. Neither structural phase transition nor Fe-site SDW ordering was
observed, consistent with the electron-doped state in the
Fe$_2$As$_2$ layers. In addition, we have identified two intrinsic
magnetic transitions in connection with the mixed valence of V. The
transition at 150 K is ascribed as an AFM ordering. The following
transition at 55 K is weakly ferromagnetic, probably due to a spin
canting process. The WFM transition renders a rare example of
coexistence of SC and WFM, which will be of great interest for the
future studies. Further experiments such as neutron diffractions are
desirable to check the spin structure of V, and to confirm that the
moment doesn't come from Fe.

\begin{acknowledgments}
This work is supported by the NSF of China (No. 90922002), National
Basic Research Program of China (No. 2007CB925001) and the
Fundamental Research Funds for the Central Universities of China
(No. 2010QNA3026). The research in Jerusalem is supported by the
Israel Science Foundation (Bikura 459/09) and by the joint
German-Israeli DIP project. FCZ acknowledges partial support from
Hong Kong RGC grant HKU 7068/09P and NSF/RGC N-HKU 726/09. The
authors would like to thank NSRL for the synchrotron radiation beam
time.
\end{acknowledgments}


\begin{thebibliography}{00}
\bibitem{hosono}Y. Kamihara, T. Watanabe, M. Hirano, and H. Hosono, J. Am.
Chem. Soc. \textbf{130}, 3296 (2008).

\bibitem{review}For a recent review, see D. C. Johnston, arXiv:1005.4392 (2010).

\bibitem{dpc}C. de la Cruz, Q. Huang, J. W. Lynn, J. Li, W. Ratcliff
II, J. L. Zarestky, H. A. Mook, G. F. Chen, J. L. Luo, N. L. Wang,
and P. Dai, Nature (London) \textbf{453}, 899 (2008).

\bibitem{bw}Q. Huang, Y. Qiu, W. Bao, M. A. Green, J. W. Lynn,
Y. C. Gasparovic, T. Wu, G. Wu, and X. H. Chen, Phys. Rev. Lett.
\textbf{101}, 257003 (2008).

\bibitem{ren-oxygen}Z. A. Ren, J. Yang, W. Lu, W. Yi, X. L. Shen, Z. C. Li, G. C.
Che, X. L. Dong, L. L. Sun, F. Zhou, and Z. X. Zhao, EPL
\textbf{82}, 57002 (2008).

\bibitem{wen-hole}H. H. Wen, G. Mu, L. Fang, H. Yang, and X. Zhu, EPL \textbf{82},
17009 (2008).

\bibitem{wang-Th}C. Wang, L. Li, S. Chi, Z. Zhu, Z. Ren, Y. Li, Y. Wang,
X. Lin, Y. Luo, S. Jiang, X. Xu, G. Cao, and Z. Xu, EPL \textbf{83},
67006 (2008).

\bibitem{Co}A. S. Sefat, A. Huq, M. A. McGuire, R. Y. Jin, B. C. Sales, D.
Mandrus, L. M. D. Cranswick, P. W. Stephens, and K. H. Stone, Phys.
Rev. B \textbf{78}, 104505 (2008); C. Wang, Y. K. Li, Z. W. Zhu, S.
Jiang, X. Lin, Y. K. Luo, S. Chi, L. J. Li, Z. Ren, M. He, H. Chen,
Y. T. Wang, Q. Tao, G. H. Cao, and Z. A. Xu, Phys. Rev. B
\textbf{79}, 054521 (2009).

\bibitem{Ni}G. H. Cao, S. Jiang, X. Lin, C. Wang, Y. K. Li, Z. Ren, Q. Tao,
C. M. Feng, J. H. Dai, Z. A. Xu, and F. C. Zhang, Phys. Rev. B
\textbf{79}, 174505 (2009).

\bibitem{ren-P}Z. Ren, Q. Tao, S. Jiang, C. M. Feng, C. Wang, J. H. Dai, G. H. Cao, and
Z. A. Xu, Phys. Rev. Lett. \textbf{102}, 137002 (2009).

\bibitem{pressure}M. S. Torikachvili, S. L. Budko, N. Ni, and P. C. Canfield,
Phys. Rev. Lett. \textbf{101}, 057006 (2008).

\bibitem{singh}D. J. Singh and M.-H. Du, Phys. Rev. Lett. \textbf{100}, 237003 (2008).
\bibitem{wnl}J. Dong, H. J. Zhang, G. Xu, Z. Li, G. Li, W. Z. Hu, D. Wu,
G. F. Chen, X. Dai, J. L. Luo, Z. Fang, and N. L. Wang, EPL
\textbf{83}, 27006 (2008).
\bibitem{mazin1}I. I. Mazin, D. J. Singh, M. D. Johannes, and M. H. Du,
Phys. Rev. Lett. \textbf{101}, 057003 (2008).
\bibitem{kuroki}K. Kuroki, S. Onari, R. Arita, H. Usui, Y. Tanaka, H. Kontani, and H. Aoki, Phys. Rev. Lett. \textbf{101},
087004 (2008).
\bibitem{ding}K. Terashima, Y. Sekiba, J. H. Bowen, K. Nakayama, T.
Kawahara, T. Sato, P. Richard, Y. M. Xu, L. J. Li, G. H. Cao, Z. A.
Xu, H. Ding, and T. Takahashi, Proc. Nat. Acad. Sci. USA
\textbf{106}, 7330 (2009).


\bibitem{wen1}X. Y. Zhu, F. Han, G. Mu, P. Cheng, B. Shen, B. Zeng, and H. H. Wen, Phys. Rev. B \textbf{79}, 220512(R)
(2009).

\bibitem{shein}I. R. Shein and A. L. Ivanovskii, J. Supercond. Nov. Magn. \textbf{22}, 613 (2009).
\bibitem{pickett}K. W. Lee and W. E. Pickett, EPL \textbf{89}, 57008 (2010).
\bibitem{wang}G. Wang, M. Zhang, L. Zheng, and Z. Yang, Phys. Rev. B \textbf{80}, 184501 (2009).
\bibitem{mazin2}I. I. Mazin, Phys. Rev. B \textbf{81}, 020507(R)
(2010).
\bibitem{nakamura}H. Nakamura and M. Machida, arXiv:1004.4741 (2010).

\bibitem{wen2}F. Han, X. Y. Zhu, G. Mu, P. Cheng, B. Shen, B. Zeng, and H. H. Wen, Sci. China G \textbf{53}, 1202 (2010).

\bibitem{tegel}M. Tegel, F. Hummel, Y. X. Su, T. Chatterji, M. Brunelli, and D. Jorhendt, EPL \textbf{89}, 37006 (2010).

\bibitem{xas-v1}J. Wong, F. W. Lytle, R. P. Messmer and D. H. Maylotte, Phys. Rev. B \textbf{30}, 5596 (1984).

\bibitem{note}The reliable factors of the refinement are $R_{wp}$=0.086 and $R_{exp}$=0.051, indicating a fairly good accuracy for the structural parameters obtained.

\bibitem{mcguire}M. A. McGuire, A. D. Christianson, A. S. Sefat, B. C. Sales, M. D. Lumsden, R. Y. Jin,
E. A. Payzant, D. Mandrus, Y. B. Luan, V. Keppens, V. Varadarajan,
J. W. Brill, R. P. Hermann, M. T. Sougrati, F. Grandjean, and G. J.
Long, Phys. Rev. B \textbf{78}, 094517 (2008).

\bibitem{luo}Y. K. Luo, Q. Tao, Y. K. Li, X. Lin, L. J. Li, G. H. Cao, Z. A. Xu, Y. Xue, H. Kaneko, A. V. Savinkov, H. Suzuki, C. Fang, and J. P. Hu, Phys. Rev. B \textbf{80}, 224511 (2009).
\bibitem{felner}I. Nowik and I. Felner, Physica C \textbf{469}, 485 (2009).
\bibitem{bvs1}I. D. Brown and D. Altermatt, Acta Cryst. \textbf{B41}, 244 (1985).
\bibitem{V/Fe}M. Tegel, T. Schmid, T. St\"{u}rzer, M. Egawa, Y. X. Su, A. Senyshyn, and D.
Johrendt, arXiv: 1008.2687.
\bibitem{RVO3}J. S. Zhou, J. B. Goodenough, J. Q. Yan, and Y. Ren, Phys. Rev. Lett. \textbf{99}, 156401 (2007).
\bibitem{gennes}P. -G. de Gennes, Phys. Rev. \textbf{118}, 141 (1960).
\bibitem{YVO3}Y. Ren, T. T. M. Palstra, D. I. Khomskii, E.
Pellegrin, A. A. Nugroho3, A. A. Menovsky, and G. A. Sawatzky,
Nature \textbf{396}, 441 (1998).

\end{thebibliography}
\end{document}